# Tin-selenium compounds at ambient and high pressures

Kien Nguyen-Cong, Joseph M. Gonzalez, Brad A. Steele, and Ivan I. Oleynik*

*Department of Physics, University of South Florida, 4202 East Fowler Ave, Tampa, FL 33620*

$Sn_xSe_y$ crystalline compounds consisting of Sn and Se atoms of varying composition are systematically investigated at pressures from 0 to 100 GPa using the first-principles evolutionary crystal structure search method based on density functional theory (DFT). All known experimental phases of SnSe and $SnSe_2$ are found without any prior input. A second order polymorphic phase transition from SnSe-$Pnma$ phase to SnSe-$Cmcm$ phase is predicted at 2.5 GPa. Initially being semiconducting, this phase becomes metallic at 7.3 GPa. Upon further increase of pressure up to 36.6 GPa, SnSe-$Cmcm$ phase is transformed to CsCl-type SnSe-$Pm\bar{3}m$ phase, which remains stable at even higher pressures. A metallic compound with different stoichiometry, $Sn_3Se_4$-$I\bar{4}3d$, is found to be thermodynamically stable from 18 GPa to 70 GPa. Known semiconductor tin diselenide $SnSe_2$-$P\bar{3}m1$ phase is found to be thermodynamically stable from ambient pressure up to 18 GPa. Initially being semiconducting, it experiences metalization at pressures above 8 GPa.

## I. INTRODUCTION

Tin selenide SnSe and tin diselenide $SnSe_2$ are layered semiconductor compounds, which are actively explored as optoelectronic, photovoltaic, and thermoelectric materials[1-7]. Similar to other metal chalcogenides, covalent bonds are formed within single layers bonded together by weak interlayer van der Waals interactions. Due to substantial interlayer distance, these layered compounds are expected to undergo substantial pressure-induced structural changes accompanied by corresponding changes in their properties. For example, semiconductor-metal transitions upon hydrostatic compression were observed in layered $MoS_2$[8] and $WS_2$[9] compounds undergoing structural changes, whereas pressure-induced metalization of $MoSe_2$[10] and $WSe_2$[11] occurred within the same crystal structures.

Compared to transition metal chalcogenides, pressure-induced phase transitions in II-VI tin selenium compounds is less understood. For example, theory[12] and experiment[13] showed that at 7 GPa $\alpha$-SnSe-$Pnma$ phase is transformed to $\beta$-SnSe-$Cmcm$ phase, the latter possessing exceptional thermoelectric properties[5,14]. This phase also appears upon increase of temperature to 800 K under ambient pressure[5]. Yet, other calculations[15,16] and experimental measurements[16,17] suggested a phase transition from $\alpha$-SnSe-$Pnma$ to $\beta'$-SnSe-$Bbmm$ at various pressures between 8 GPa and 18 GPa. There is a confusion in the literature concerning usage of space group names, $Bbmm$ and $Cmcm$, both belonging to the the same space group number 63. Initially, $Cmcm$ and $Bbmm$ phases were proposed to have tetragonal and orthorhombic lattices, respectively[17]. However, a later characterization[18] showed that they are both orthorhombic. Therefore, we consider $\beta$-SnSe-$Cmcm$ and $\beta'$-SnSe-$Bbmm$ phases to be identical and will use the label for this structure, $\beta$-SnSe-$Cmcm$, in this work.

Uncertainty in pressure-induced metalization of SnSe also exists: the experiment by Agarwal *et al*[19] found that the semiconductor-metal transition occurs at 6.5 GPa whereas the calculations of Yan *et al*[20] reported the met-alization pressure of 12 GPa. Upon further increase of pressure up to 27 GPa $\beta$-SnSe-$Cmcm$ phase transforms to CsCl-type $Pm\bar{3}m$ phase, which is superconducting at $T < T_c = 4.5$ K[16,21].

Effects of pressure on structure and properties of $SnSe_2$ are also not well understood. At ambient conditions, $SnSe_2$ crystallizes in $P\bar{3}m1$ phase. There are no reports of pressure-induced phase transitions thus far. Although a few studies[22-24] showed reduction of the band gap of $SnSe_2$ upon compression at low pressures, pressure-induced metalization or appearance of new metallic phases has not been reported yet.

Motivated by the above-mentioned outstanding problems, we perform a systematic study of structural, electronic and vibrational properties of the broad class of tin selenium compounds with varying stoichiometry at ambient conditions and under compression up to 100 GPa. Evolutionary crystal structure searches based on first-principles density functional theory (DFT) are performed at various pressures with the goal to discover all stable tin/selenium compounds within the pressure range of 0-100 GPa. Once individual $Sn_xSe_y$ compounds are found, they are characterized by calculating their pressure-dependent structural, electronic and vibrational properties.

## II. COMPUTATIONAL METHODS

The evolutionary crystal structure search of $Sn_xSe_y$ compounds with variable stoichiometry is performed at 0 GPa, 15GPa, 30 GPa, 60 GPa and 100 GPa using USPEX code[25]. Initially, random structures are generated by USPEX and are relaxed using first principles DFT Vienna *Ab initio* Simulation Package (VASP)[26,27]. Then, the relaxed structures are ranked based on their formation enthalpy, which is calculated using the ground state crystal structures of Sn and Se at corresponding pressures. At the next generation, USPEX produces the set of offspring structures by applying random, heredity and mutation operators, followed by structure optimization by VASP, ranking optimized structures by the forma-tion enthalpy, and constructing of new generation. These steps are repeated until the lowest enthalpy structures for each stoichiometry remain for at least ten generations. Depending on the targeted pressure, $Pnma$, $P6/mmm$, $\beta$ or bcc phases are the lowest enthalpy phases of Sn, whereas $\alpha$ or $\beta$ Po-type phases are the lowest enthalpy phases of Se at corresponding pressures.

During the structure search, first principles calculations are performed using PBE generalized gradient approximation (GGA) to DFT[28] plane-wave energy cut-off of 500 eV, k-point spacing in the Brillouin zone of 0.06 Å$^{-1}$, and projector augmented wave potentials (PAW)[29]. Such parameters allowed to achieve the formation enthalpy accuracy better than 10 meV/atom. The constant pressure geometry optimization is achieved using the force convergence criterion 0.07 eV/Å. Once the search is complete, the lowest enthalpy structures are reoptimized using fine VASP parameters with energy cutoff 700 eV, k-point density 0.03 Å$^{-1}$ and force tolerance less than 0.01 eV/Å to achieve the convergence of formation enthalpy better than 1 meV/atom.

Van der Waals (vdW) interactions play an important role in layered $Sn_xS_y$ materials. Therefore, to properly account for vdW interactions, we evaluated performance of several correction methods, which are currently used to address the deficiencies of standard DFT calculations. They include empirical Grimme-D2[30] and semi-empirical Tkatchenko-Scheffler (TS)[31] methods, vdW functional by Langreth and Lundqvist (DF2)[32,33], and recently developed revised-vv10 (rvv10) vdW functional[34]. In Tables I and II lattice parameters and cell volumes of $\alpha$-SnSe-$Pnma$ and SnSe$_2$-$P\bar{3}m1$ phases at 0 GPa calculated using the methods listed above are compared to those calculated using standard PBE GGA as well as those measured in experiment. Based on these tests it is found that Grimme-D2 method provides best agreement with experiment at both 0 GPa and high pressure conditions (e.g. the lattice parameter of cubic $Pm\bar{3}m$-SnSe at 50.1 GPa is 3.19 Å, which is in good agreement with experimental value of 3.18 Å[16]).

To achieve an accurate description of electronic properties of semiconducting phases of tin/selenium compounds including band gaps, screened hybrid functional HSE06[35,36] is employed. In this case, smaller energy cutoff of 400 eV is used to achieve satisfactory accuracy at a reasonable computational cost.

Phonon dispersion of selected structures is calculated using frozen phonon technique[39] as implemented in Phonopy code[40] using 2x2x2 supercells. Force calculation used to determine phonon frequencies is found to be sensitive to accurate atomic positions, therefore, the geometry optimization is done using much finer, 0.001 eV/Å, force convergence criterion.

## III. RESULTS AND DISCUSSION

### A. Convex hulls and phase diagram

In order to determine thermodynamically stable structures in an interval of pressures from 0 to 100 GPa, the convex hull of tin selenide $Sn_xSe_y$ compounds is constructed at several target pressures. The convex hull connects the formation enthalpy/stoichiometry points of the lowest formation enthalpy structures: if a structure with a particular stoichiometry is on the convex hull, then it has a negative heat of formation and is thermodynamically stable, i.e. does not spontaneously decompose on elemental Sn and Se or any other $Sn_xSe_y$ compounds. The $Sn_xSe_y$ convex hulls at 0, 15, 30, 60 and 100 GPa are represented in Fig. 1.

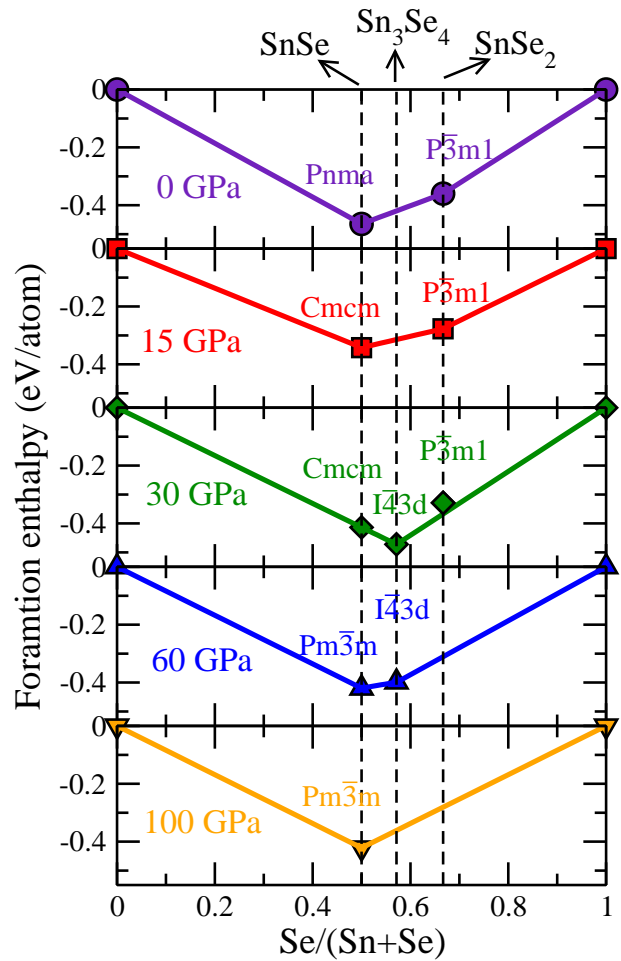

Figure 1. Convex hull of $Sn_xSe_y$ at 0, 15, 30, 60 and 100 GPa.

Our crystal structure search is able to predict all known phases of tin selenide compounds at ambient and high pressures, thus demonstrating the robustness of USPEX structure prediction. Specifically, $\alpha$-SnSe-$Pnma$ and SnSe$_2$-$P\bar{3}m1$ phases are predicted to be thermodynamically stable at 0 GPa. In addition to the $\alpha$ phase of SnSe, the search finds two other metastable phases: $\beta$-SnSe-$Cmcm$ and rocksalt-type SnSe-$Fm\bar{3}m$, which are respectively 2 meV and 22 meV higher that the $Pnma$ phase. At 15 GPa, the stoichiometries of stable compounds (i.e. SnSe and SnSe$_2$) do not change, but the $\alpha$-SnSe-$Pnma$ to $\beta$-SnSe-$Cmcm$ phase transition occurs at some pressure below 15 GPa (see discussion below), the latter phase being thermodynamically stable up to 36 GPa.

Table I. The calculated lattice parameters and cell volume of $\alpha$-SnSe-$Pnma$ phase at 0 GPa compared to experiment. Percentage in parentheses indicates the difference between experiment and theory.

| Lattice | Theor. (PBE)[12] | Exp.[5] | PBE | PBE+D2 | PBE+TS | DF2 | Rvv10 |
|---|---|---|---|---|---|---|---|
| a (Å) | 11.69 (+1.04%) | 11.57 | 11.76(+1.64%) | 11.63(+0.52%) | 11.64(+0.61%) | 12.50(+8.04%) | 12.39(+7.09%) |
| b (Å) | 4.23 (+0.95%) | 4.19 | 4.20(+0.23%) | 4.20(+0.24%) | 4.20(+0.24%) | 4.334(+3.44%) | 4.28(+2.15%) |
| c (Å) | 5.52 (+23.77%) | 4.46 | 4.56(+2.42%) | 4.36(-2.42%) | 4.45(+0.22%) | 4.86(+8.97%) | 4.35(-2.47%) |
| Vol (Å$^3$) | 272.95(+26.24%) | 216.21 | 225.21(+4.16%) | 212.63(-1.66%) | 218.21(+0.93%) | 263.34(+21.80%) | 231.43(+7.04)% |

Table II. The calculated lattice parameters and cell volume of SnSe$_2$-$P\bar{3}m1$ phase at 0 GPa compared to experiment. Percentage in parentheses indicates the difference between experiment and theory.

| Lattice | Theor. (PBE)[37] | Exp.[38] | PBE | PBE+D2 | PBE+TS | DF2 | Rvv10 |
|---|---|---|---|---|---|---|---|
| a (Å) | 3.88 (+1.83%) | 3.81 | 3.87(+1.57%) | 3.83(+0.52%) | 3.87(+1.57%) | 4.04(+6.04%) | 3.93(+3.15%) |
| c (Å) | 6.74 (+9.77%) | 6.14 | 6.96(+13.36%) | 6.16(+0.33%) | 6.27(+2.12%) | 6.33(+3.09%) | 6.17(+0.49%) |
| Vol (Å$^3$) | 87.87 (+13.83%) | 77.19 | 94.38(+22.27%) | 78.31(+1.45%) | 81.39(+5.44%) | 89.45(+15.88%) | 82.39(6.73%) |

At 30 GPa, a new compound with a stoichiometry Sn$_3$Se$_4$ appears at the convex hull, see Fig. 1. It has $I\bar{4}3d$ symmetry and displays the lowest formation enthalpy compared to another thermodynamically stable $\beta$-SnSe-$Cmcm$ phase at 30 GPa. The tin diselenide crystal SnSe$_2$-$P\bar{3}m1$ is not on the convex hull at 30 GPa and is therefore energetically more favorably for it to decompose into pure Se in $\beta$ Po-type phase and Sn$_3$Se$_4$-$I\bar{4}3d$. The existence of Sn$_3$Se$_4$-$I\bar{4}3d$ has recently been reported in Ref. [41].

At 60 GPa, there exist only two thermodynamically stable compounds with stoichiometries SnSe and Sn$_3$Se$_4$. In contrast to the Sn$_3$Se$_4$ stoichiometry that remains in the same Sn$_3$Se$_4$-$I\bar{4}3d$ phase as that at 30 GPa, the SnSe crystal undergoes a phase transition to the cubic CsCl-type $Pm\bar{3}m$ phase (SnSe-$Pm\bar{3}m$), which possesses the lower formation enthalpy than Sn$_3$Se$_4$-$I\bar{4}3d$ phase, see Fig. 1. As pressure increases to 100 GPa, Sn$_3$Se$_4$-$I\bar{4}3d$ phase is no longer thermodynamically stable, resulting in the only SnSe-$Pm\bar{3}m$ compound on the convex hull. In addition to pressures of 0, 15, 30, 60 and 100 GPa, the convex hulls are constructed at intermediate pressures to build the phase diagram for all stable compounds of tin and selenium in the pressure range from 0 to 100 GPa, see Fig. 2. The phase diagram displays the structures that are both thermodynamically and dynamically stable. The first criterion is judged by the presence of a particular structure on the convex hull, see Fig. 1. The second criterion of dynamical stability requires an absence of imaginary frequency modes in the phonon spectrum for a particular structure, see below. The corresponding phase transitions are discussed below.

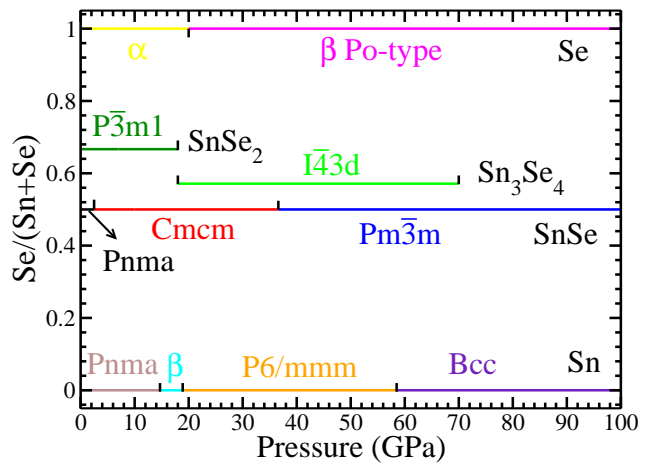

Figure 2. The pressure-composition phase diagram of Sn$_x$Se$_y$ compounds from 0 to 100 GPa.

### B. Phase transitions and pressure-dependent properties of SnSe compounds

Among all three types of tin/selenium compounds, SnSe exhibits the richest polymorphism as at least three phases exist at pressures up to 100 GPa: $\alpha$-SnSe-$Pnma$, $\beta$-SnSe-$Cmcm$, and SnSe-$Pm\bar{3}m$, see the phase diagram in Fig. 2. The crystal structures of two nontrivial layered crystals, $\alpha$-SnSe-$Pnma$ and $\beta$-SnSe-$Cmcm$, are shown in the Fig. 3(a), whereas SnSe-$Pm\bar{3}m$ phase possesses simpler CsCl structure, is not shown there.

At ambient conditions the SnSe crystal is in $\alpha$-SnSe-$Pnma$ phase. Upon compression it transforms into a topologically similar phase $\beta$-SnSe-$Cmcm$, the latter distinguishes itself by symmetric positions of Sn atoms; one of them is marked with a red circle in of Fig. 3(a). The z coordinate of this atom along the crystallographic c axis serves as an order parameter for the phase transitions and its pressure evolution is plotted in Fig. 3(b) as one phase $\alpha$-SnSe-$Pnma$ transforms to another, $\beta$-SnSe-$Cmcm$ phase. According to the evolution of the order parameter, the phase transition is complete at 2.5 GPa as the fractional z-coordinate of a Sn atom converges to 0.5, see Fig. 3(b).

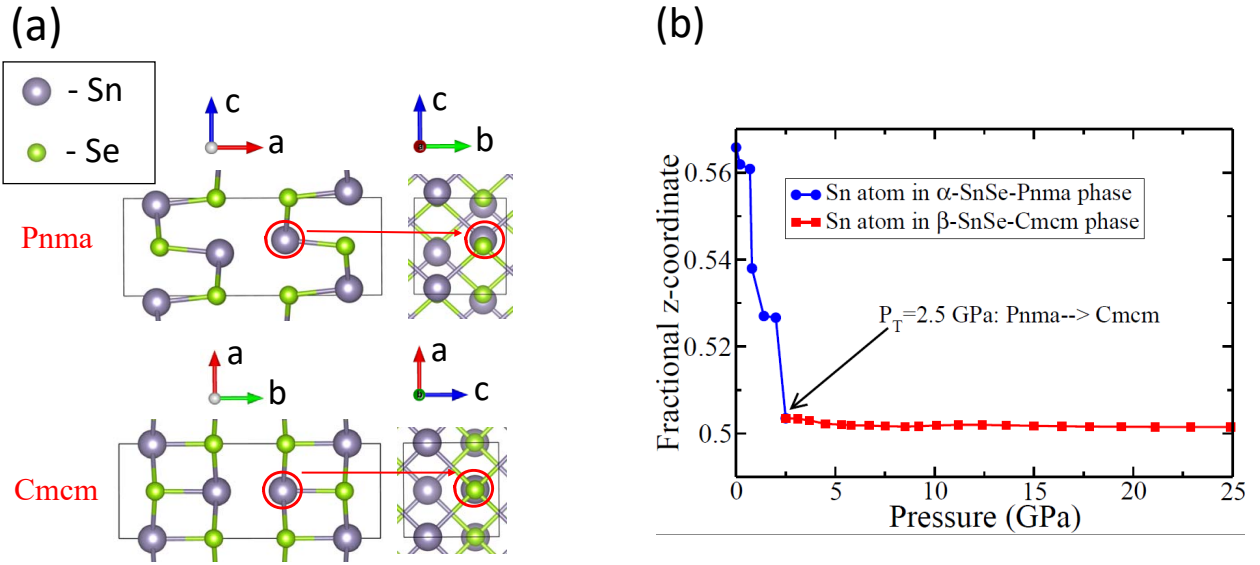

Figure 3. (a) Crystal structures (side and top views) of two SnSe phases: $\alpha$-SnSe-$Pnma$ and $\beta$-SnSe-$Cmcm$. Although both phases are topologically similar, their crystallographic axes are labeled differently according to International Tables for X-ray Crystallography: $a$ is the longest dimension of the $Pnma$ cell, whereas $c$ is longest unit cell dimension of $Cmcm$ crystal. (b) Fractional z coordinate of the red-circled Sn atom in Fig. 3(a) as a function of pressure, which is used as an order parameter to differentiate between two phases: $\alpha$-SnSe-$Pnma$ and $\beta$-SnSe-$Cmcm$. The phase transition occurs at 2.5 GPa

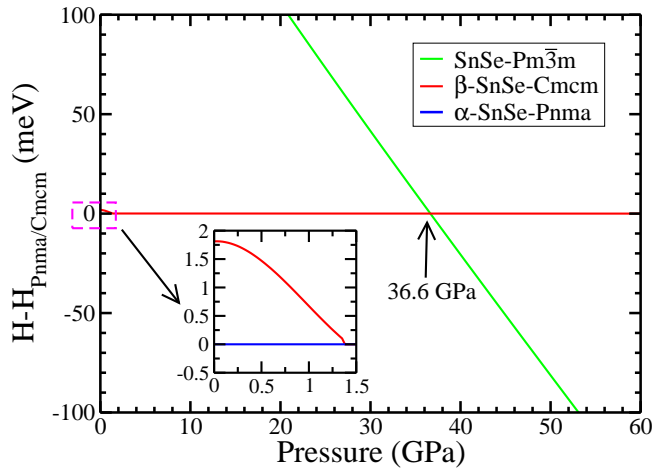

Figure 4. Enthalpy vs pressure for $\alpha$-SnSe-$Pnma$, $\beta$-SnSe-$Cmcm$, and SnSe-$Pm\bar{3}m$ phases. Enthalpy the of $Pnma$ phase (before the phase transition) and the $Cmcm$ phase (after the phase transition) is used as a reference enthalpy.

The evolution of formation enthalpy $H$ of all three phases of SnSe upon compression is shown in Fig. 4. The enthalpy difference between the 0 GPa phase $\alpha$-SnSe-$Pnma$ and $\beta$-SnSe-$Cmcm$ phase is small (less than 2 meV/atom at 0 GPa) and becomes substantially below the accuracy of our DFT calculations (1 meV/atom) at the phase transition, see inset in Fig. 4. Therefore, the transition pressure can not be reliably resolved from formation enthalpy differences between the phases. However, the geometrical order parameter (the fractional z coordinate of the Sn atom) is more reliable indicator of the transition that occurs at 2.5 GPa.

The pressure vs volume equation of state (EOS) for all three phases of SnSe is shown in Fig. 5. Based on the behavior of EOS at the phase transition points it is possible to make a conclusion about the nature of the transition. As there is no discontinuity between $\alpha$-SnSe-$Pnma$ and $\beta$-SnSe-$Cmcm$ phases at 2.5 GPa, we make a conclusion that it is the second order phase transition. Correspondingly, $\beta$-SnSe-$Cmcm$ to SnSe-$Pm\bar{3}m$ phase transition at 36.6 GPa displays a discontinuity in $P(V)$ EOS, which characterizes this transition as the first order, see Fig. 5.

The dynamical stability of various thermodynamically stable phases is checked by calculating the phonon spectra of these crystals to make sure that there are no imaginary frequency modes. In particular, the $\beta$-SnSe-$Cmcm$ phase is unstable at 0 GPa and 0 K as it carries imaginary frequencies near $\Gamma$ and Y $q$ points, see Fig. 6(a-b). However, it becomes dynamically stable above 2 GPa, see Fig. 6(c), which confirms that this phase is the ground state phase at this pressure. The observed stability of $\beta$-SnSe-$Cmcm$ phase at ambient pressure and elevated temperatures can be explained by a substantial contribution of anharmonic effects[42–44]. The predicted pressure 2.5 GPa of α-SnSe-$Pnma$ to β-SnSe-$Cmcm$ phase transition is below the values obtained in previous experiment[17] and theory[45]. The disagreement with the latter work of Yu et al comes from the fact that although vdW interactions play a crucial role in these layered compounds, they have not been included by Yu et al[45], whereas they are properly accounted for in the present work. Experimental pressure in Ref.[17] might be affected by the polycrystalline nature of the samples.

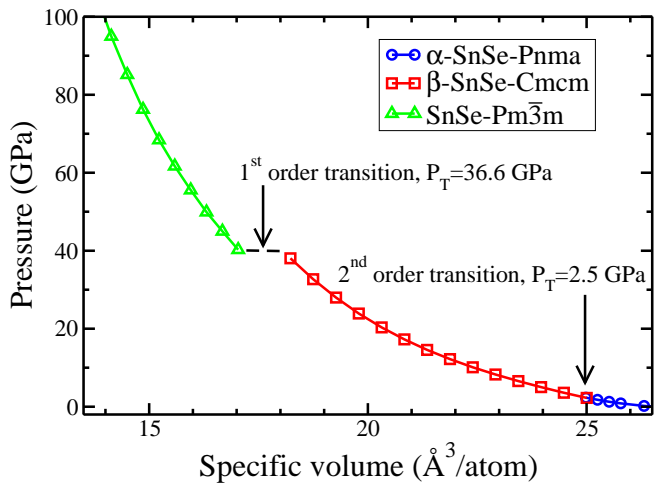

Figure 5. Equation of state of three SnSe phases: α-SnSe-$Pnma$, β-SnSe-$Cmcm$, and SnSe-$Pm\bar{3}m$.

To complete the discussion of SnSe phases, the pressure dependence of their electronic properties is considered. At 0 GPa, our HSE DFT calculations demonstrate that α-SnSe-$Pnma$ phase is an indirect band gap semiconductor with $E_g = 0.96$ eV, see Fig. 7(a), which is in a good agreement with previous calculations (0.9eV)[46] and experimental measurements (0.95eV)[47]. Upon compression, the calculated band gap of α-SnSe-$Pnma$ phase is reduced monotonically and reaches 0.56 eV at 1.9 GPa, see Fig. 7(c). At the α-SnSe-$Pnma$ to β-SnSe-$Cmcm$ phase transition point of 2.5 GPa, $E_g(P)$ does not experience any discontinuity which is consistent with the second order nature of the phase transition. Upon further compression of the β-SnSe-$Cmcm$ phase, $E_g$ monotonically decreases and becomes zero at around 7.3 GPa, see Fig. 7(b). The predicted semiconductor/metal transition pressure of 7.3 GPa is in good agreement with experimental pressure of metalization at 6.5 GPa[19].

The band structure of two SnSe phases is displayed in Fig. 7(a): α-SnSe-$Pnma$ phase at 0 GPa, and Fig. 7(b): β-SnSe-$Cmcm$ phase at 6.5 GPa. In these two cases, both valence band maximum (VBM) and conduction band minimum (CBM) are located at different k-points along Γ-Z direction of the Brillouin zone. As pressure increases, valence and conduction bands move toward each other, while positions of VBM and CBM do not change. The metalization that occurs at 7.3 GPa in β-SnSe-$Cmcm$ phase is due to indirect overlap of VBM and CBM.

## C. Crystal structure and properties of Sn₃Se₄ -$I\bar{4}3d$ compound

As discussed above, our structure search discovered a new compound with stoichiometry Sn₃Se₄ that is stable in the pressure range 18-70 GPa, see Fig. 2, and has a primitive rhombohedral unit cell containing two formula units (f.u.), see Fig. 8. The Wyckoff positions of the atoms in the cubic conventional cell containing 4 f.u. are listed in Table III. Its 3-dimensional structure consists of corner sharing SnSe₈ units connected in the network where each Se atom is shared by every three Sn atoms and each Sn atom is surrounded by eight Se atoms. At 30 GPa, the first and the second nearest neighbor Sn-Se distances are 2.677 Å and 2.863 Å. In comparison, both Sn and Se atoms of β-SnSe-$Cmcm$ crystal have coordination numbers of 8 and first and second nearest neighbor distances are 2.604 Å and 2.718 Å respectively at the same pressure of 30 GPa. In SnSe₂-$P\bar{3}m1$ crystal, the first nearest-neighbor Sn-Se distance is 2.604 Å, and Sn (Se) has coordination number of 6 (3).

Table III. Wyckoff positions of atoms in Sn₃Se₄-$I\bar{4}3d$ conventional unit cell with conventional lattice parameter a=7.9943 Å at 30 GPa.

| Wyckoff position | x | y | z |
|---|---|---|---|
| Sn (12a) | -0.1250 | -0.5000 | 0.7500 |
| Se (16c) | 0.1779 | -0.66779 | 0.6779 |

Our phonon calculations demonstrate that Sn₃Se₄-$I\bar{4}3d$ is also dynamically stable in the pressure range 18-70 GPa, see Fig. 9. In particular, it has imaginary frequencies at 10 GPa, but it is dynamically stabilized at high pressures as phonon frequencies increase with pressure, see, for example, an absence of imaginary frequencies in the phonon spectra at 30 and 100 GPa in Figs. 9 (b) and (c). Although Sn₃Se₄-$I\bar{4}3d$ is not at the convex hull at 100 GPa, it does not have any imaginary frequency modes, therefore, can be considered as metastable at this pressure.

The electronic structure calculations of Sn₃Se₄-$I\bar{4}3d$ crystal demonstrates that it is a metal in the entire range of pressures of its stability, see its band structure and density of states at 30 GPa in Fig. 10. Partial density of states (PDOS) demonstrates that the major contribution to the states at the Fermi energy are from $4s$ and $4p$ states of Se and $5p$ states of Sn.

## D. Pressure-induced semiconductor-metal transition in SnSe₂

Our calculations of thermodynamic and dynamic stabilities of SnSe₂-$P\bar{3}m1$ demonstrate that it is stable from ambient pressure up to 18 GPa, see Figs. 1 and 2. At higher pressures above 18 GPa it becomes metastable and then dynamically unstable at 41 GPa. It is known that at ambient pressure $SnSe_2$ phase is an indirect band gap semiconductor[48]. Our HSE calculation of the band structure at 0 GPa, shown in Fig. 11(a), demonstrate that it has an indirect band gap of 1.09 eV, which is con-sistent with previous experimental studies of Domingo (0.97 eV)[49] and Manou (1.06 eV)[50] and HSE calculations (1.07 eV)[48]. The band gap is between the VBM at a point between $\Gamma$ and K of BZ, and CBM at a point on the M-L segment of BZ, see Fig. 11(a). Upon compression, band gap reduces linearly with pressure, see 11(c). This decrease is due to a spread of the conduction band downwards to the Fermi level upon increase of pressure. Simultaneously, CBM experiences a gradual shift from the point between M and L to a valley minimum at K, for example see band structure at 5.6 GPa in Fig. 11(b). At about 8 GPa, the band gap of $SnSe_2$-$P\bar{3}m1$ is closed by an indirect overlap of VBM and CBM. Similar to SnSe phase, the semiconductor-metal transition occurs at much lower pressures that those for transition metal dichalcogenides.

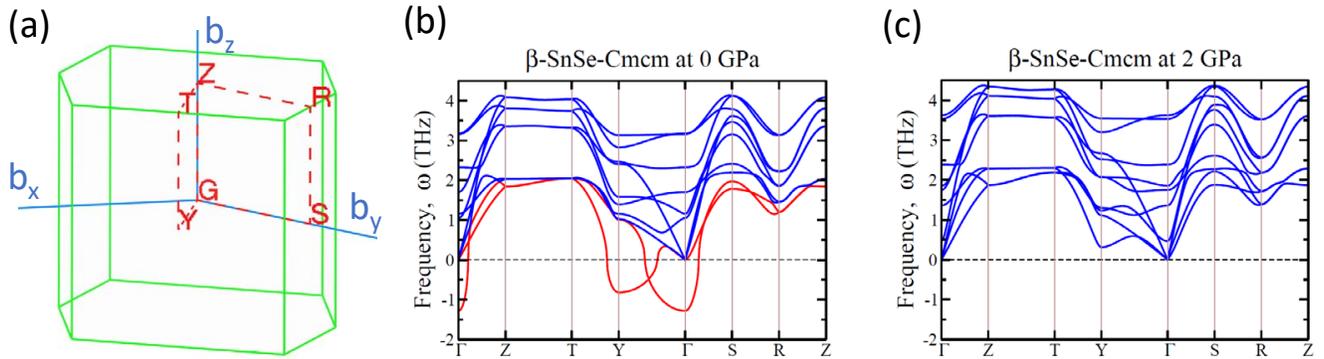

Figure 6. Phonon band structure of the $\beta$-SnSe-$Cmcm$ phase: (a) Sketch of the Brillouin zone (green solid) corresponding to the primitive direct lattice; (b) and (c) are the phonon dispersion curves at 0 GPa and 2 GPa respectively along high symmetry directions of the Brillouin zone. The $\omega(q)$ bands carrying imaginary frequencies are highlighted by red color. As there are no imaginary frequencies at 2 GPa, the $\beta$-SnSe-$Cmcm$ phase is dynamically stable above this pressure.

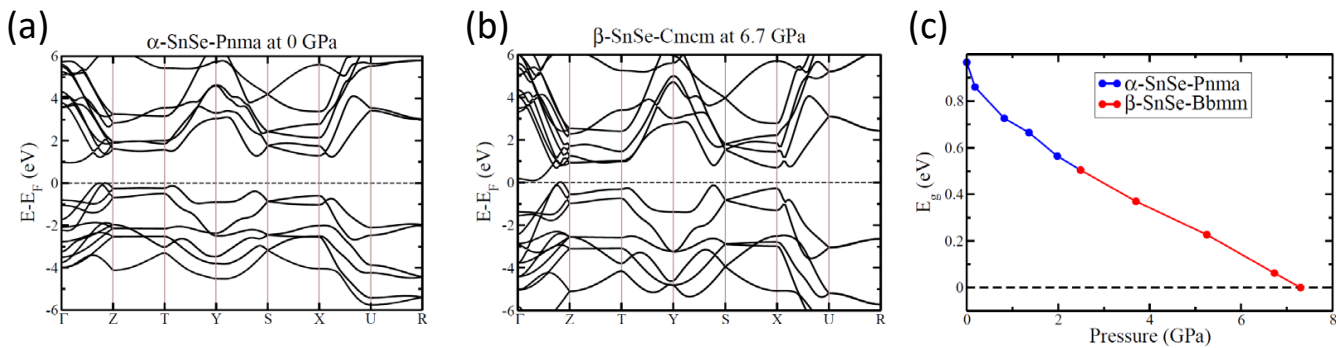

Figure 7. Pressure dependence of electronic properties of $\alpha$-SnSe-$Pnma$ and $\beta$-SnSe-$Cmcm$ phases: (a) band structure of $\alpha$-SnSe-$Pnma$ phase at 0 GPa; (b) band structure of $\beta$-SnSe-$Cmcm$ phase at 6.7 GPa; (c) band gap as a function of pressure for these two phases.

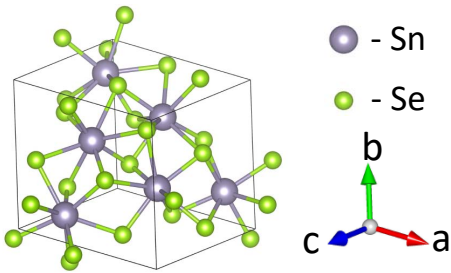

Figure 8. Crystal structure of $Sn_3Se_4$-$I\bar{4}3d$ compound consisting of 3D network of $SnSe_8$ structural units.

## IV. CONCLUSIONS

Using first-principles evolutionary crystal search, $Sn_xS_y$ compounds of variable stoichiometry are system-atically investigated at ambient conditions and pressures up to 100 GPa. Their thermodynamic and dynamic stabilities are calculated and the pressure-dependent phase diagram is constructed. Various known phases such as SnSe₂-$P\bar{3}m1$, all three SnSe phases, $\alpha$-SnSe-$Pnma$, $\beta$-SnSe-$Cmcm$ and SnSe-$Pm\bar{3}m$ are found without any prior input to the search. A second order phase transition from $\alpha$-SnSe-$Pnma$ phase to $\beta$-SnSe-$Cmcm$ phase is predicted to occur at 2.5 GPa, the latter phase becoming metallic at about 7.3 GPa. Upon further compression, a first order phase transition $\beta$-SnSe-$Cmcm$ phase to SnSe-$Pm\bar{3}m$ phase takes place at 36.6 GPa. In addition, a new metallic compound Sn₃Se₄-$I\bar{4}3d$ is found to be stable between 18 GPa and 70 GPa. Well-known ambient pressure semiconductor, tin diselenide SnSe₂-$P\bar{3}m1$, metalizes at 8 GPa, and becomes thermodynamically unstable above 18 GPa. In our recent study of similar tin-sulfur compounds, a complete phase diagram of Sn$_x$S$_y$ system has also been constructed. Although most of the phases and stoichiometries in both Sn$_x$S$_y$ and Sn$_x$Se$_y$ systems are the same, the actual transition pressures are found to be different. Such differences exemplified in complete phase diagrams are of importance for future experimental investigations of these compounds.

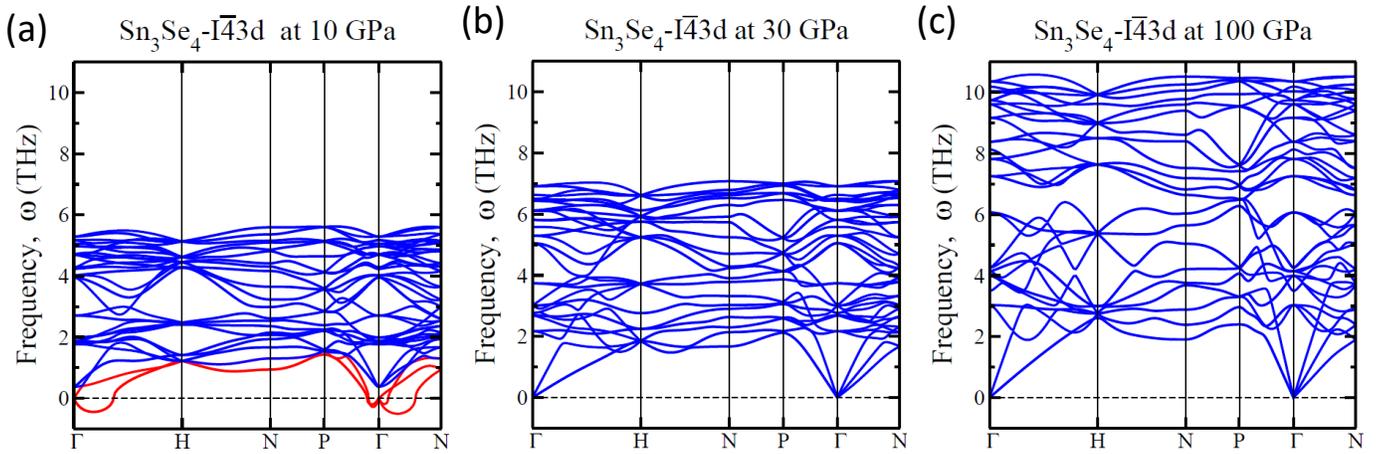

Figure 9. Phonon dispersion $\omega(q)$ of Sn₃Se₄-$I\bar{4}3d$ crystal at (a) 10 GPa, (b) 30 GPa, and (c) 100 GPa. The imaginary frequency bands at 10 GPa are highlighted by red color.

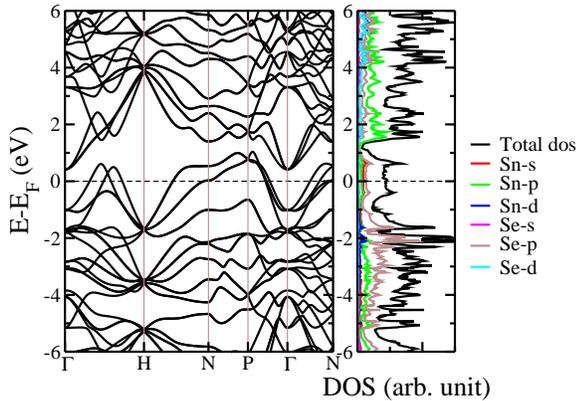

Figure 10. Band structure along high symmetry points and partial density states of $I\bar{4}3d$-Sn₃Se₄ at 30 GPa. The Fermi level is shifted to 0 eV.

### ACKNOWLEDGMENTS

We would like to acknowledge financial support by Defense Threat Reduction Agency, grant HDTRA1-12-1-0023 and Army Research Laboratory through Cooperative Agreement W911NF-16-2-0022. Simulations were performed using the NSF XSEDE facilities (grant No. TG-MCA08X040), DOE BNL CFN computational user facility, and USF Research Computing Cluster supported by NSF (grant No. CHE-1531590).

* oleynik@usf.edu
[1] W. J. Baumgardner, J. J. Choi, Y.-F. Lim, and T. Hanrath, Journal of the American Chemical Society **132**, 9519 (2010).
[2] N. Mathews, Solar Energy **86**, 1010 (2012).
[3] F. Liu, J. Zhu, Y. Xu, L. Zhou, Y. Li, L. Hu, J. Yao, and S. Dai, Chem. Commun. **51**, 8108 (2015).
[4] X. Zhou, L. Gan, W. Tian, Q. Zhang, S. Jin, H. Li, Y. Bando, D. Golberg, and T. Zhai, Advanced Materials **27**, 8035 (2015).
[5] L.-D. Zhao, S.-H. Lo, Y. Zhang, H. Sun, G. Tan, C. Uher, C. Wolverton, V. P. Dravid, and M. G. Kanatzidis, Nature **508**, 373 (2014).

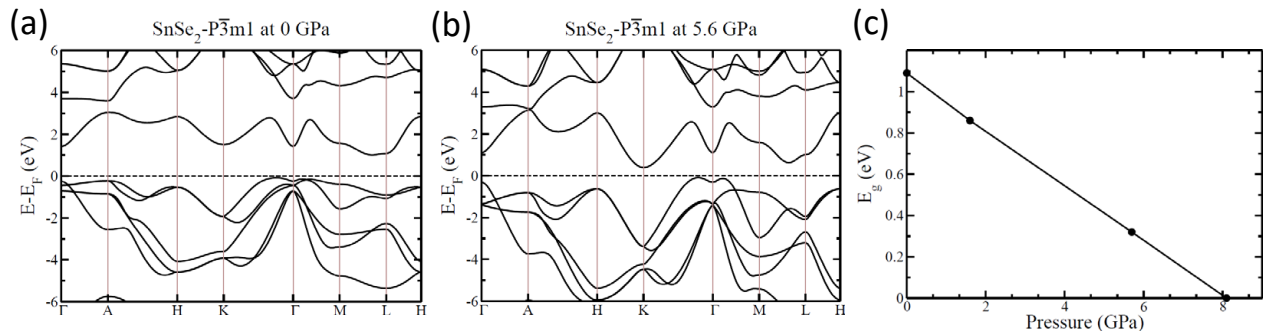

Figure 11. Band structure along high symmetry directions of the Brillouin zone at (a) 0 GPa; (b) at 5.6 GPa. (c) Pressure dependence of band gap.

[6] X. Guan, P. Lu, L. Wu, L. Han, G. Liu, Y. Song, and S. Wang, Journal of Alloys and Compounds **643**, 116 (2015).

[7] A. Dewandre, O. Hellman, S. Bhattacharya, A. H. Romero, G. K. H. Madsen, and M. J. Verstraete, Physical Review Letters **117**, 276601 (2016).

[8] A. P. Nayak, S. Bhattacharyya, J. Zhu, J. Liu, X. Wu, T. Pandey, C. Jin, A. K. Singh, D. Akinwande, and J.-F. Lin, Nature Communications **5** (2014), 10.1038/ncomms4731.

[9] A. P. Nayak, Z. Yuan, B. Cao, J. Liu, J. Wu, S. T. Moran, T. Li, D. Akinwande, C. Jin, and J.-F. Lin, ACS Nano **9**, 9117 (2015).

[10] Z. Zhao, H. Zhang, H. Yuan, S. Wang, Y. Lin, Q. Zeng, G. Xu, Z. Liu, G. K. Solanki, K. D. Patel, Y. Cui, H. Y. Hwang, and W. L. Mao, Nature Communications **6**, 7312 (2015).

[11] B. Liu, Y. Han, C. Gao, Y. Ma, G. Peng, B. Wu, C. Liu, Y. Wang, T. Hu, X. Cui, W. Ren, Y. Li, N. Su, H. Liu, and G. Zou, The Journal of Physical Chemistry C **114**, 14251 (2010).

[12] S. Alptekin, Journal of Molecular Modeling **17**, 2989 (2011).

[13] J. Z. Jian Zhang, H. Zhu, X. Wu, H. Cui, D. Li, J. Jiang, C. Gao, Q. Wang, and Q. Cui, Nanoscale **7**, 10807 (2015).

[14] H. Zhang and D. V. Talapin, Angewandte Chemie International Edition **53**, 9126 (2014).

[15] Y. Zhang, S. Hao, L.-D. Zhao, C. Wolverton, and Z. Zeng, J. Mater. Chem. A **4**, 12073 (2016).

[16] X. Chen, P. Lu, X. Wang, Y. Zhou, C. An, Y. Zhou, H. Gao, Z. Guo, C. Park, B. Hou, and K. Peng, arXiv:1608.06763 , 1 (2016), arXiv:1608.06763.

[17] I. Loa, R. J. Husband, R. A. Downie, S. R. Popuri, and J.-W. G. Bos, Journal of Physics: Condensed Matter **27**, 072202 (2015).

[18] M. Sist, J. Zhang, and B. Brummerstedt Iversen, Acta Crystallographica Section B Structural Science, Crystal Engineering and Materials **72**, 310 (2016).

[19] A. Agarwal, P. H. Trivedi, and D. Lakshminarayana, Crystal Research and Technology **40**, 789 (2005).

[20] J. Yan, F. Ke, C. Liu, L. Wang, Q. Wang, J. Zhang, G. Li, Y. Han, Y. Ma, and C. Gao, Phys. Chem. Chem. Phys. **18**, 5012 (2016).

[21] Y. A. Timofeev, B. V. Vinogradov, and V. B. Begoulev, Physics of the Solid State **39**, 207 (1997).

[22] A. I. Likhter, E. G. Pel, and S. I. Prysyazhnyuk, Physica Status Solidi (a) **14**, 265 (1972).

[23] M. J. Powell and A. J. Grant, Il Nuovo Cimento B Series 11 **38**, 486 (1977).

[24] S. V. Bhatt, M. Deshpande, V. Sathe, and S. Chaki, Solid State Communications **201**, 54 (2015).

[25] A. R. Oganov and C. W. Glass, The Journal of Chemical Physics **124**, 244704 (2006).

[26] G. Kresse and J. Furthmüller, Computational Materials Science **6**, 15 (1996).

[27] G. Kresse and J. Furthmüller, Physical Review B **54**, 11169 (1996).

[28] J. P. Perdew, K. Burke, and M. Ernzerhof, Physical Review Letters **77**, 3865 (1996).

[29] P. E. Blöchl, Physical Review B **50**, 17953 (1994).

[30] S. Grimme, Journal of Computational Chemistry **27**, 1787 (2006).

[31] A. Tkatchenko and M. Scheffler, Physical Review Letters **102**, 073005 (2009).

[32] M. Dion, H. Rydberg, E. Schröder, D. C. Langreth, and B. I. Lundqvist, Physical Review Letters **95**, 109902 (2005).

[33] J. Klimeš, D. R. Bowler, and A. Michaelides, Journal of Physics: Condensed Matter **22**, 022201 (2010).

[34] R. Sabatini, T. Gorni, and S. de Gironcoli, Physical Review B **87**, 041108 (2013).

[35] J. Heyd, G. E. Scuseria, and M. Ernzerhof, The Journal of Chemical Physics **118**, 8207 (2003).

[36] J. Heyd, G. E. Scuseria, and M. Ernzerhof, The Journal of Chemical Physics **124**, 219906 (2006).

[37] C. Xia, J. An, S. Wei, Y. Jia, and Q. Zhang, Computational Materials Science **95**, 712 (2014).

[38] B. Pałosz and E. Salje, Journal of Applied Crystallography **22**, 622 (1989).

[39] K. Parlinski, Z. Q. Li, and Y. Kawazoe, Physical Review Letters **78**, 4063 (1997).

[40] A. Togo and I. Tanaka, Scripta Materialia **108**, 1 (2015).

[41] H. Yu, W. Lao, L. Wang, K. Li, and Y. Chen, Physical Review Letters **118**, 137002 (2017).

[42] C. W. Li, J. Hong, A. F. May, D. Bansal, S. Chi, T. Hong, G. Ehlers, and O. Delaire, Nature Physics **11**, 1063 (2015).

[43] D. Bansal, J. Hong, C. W. Li, A. F. May, W. Porter, M. Y. Hu, D. L. Abernathy, and O. Delaire, Physical Review B **94**, 054307 (2016).

[44] J. M. Skelton, L. A. Burton, S. C. Parker, A. Walsh, C.-E. Kim, A. Soon, J. Buckeridge, A. A. Sokol, C. R. A.

Catlow, A. Togo, and I. Tanaka, Physical Review Letters **117**, 075502 (2016).

[45] H. Yu, S. Dai, and Y. Chen, Scientific Reports **6**, 26193 (2016).

[46] I. Lefebvre, M. A. Szymanski, J. Olivier-Fourcade, and J. C. Jumas, Physical Review B **58**, 1896 (1998).

[47] P. Pramanik and S. Bhattacharya, Journal of Materials Science Letters **7**, 1305 (1988).

[48] J. M. Gonzalez and I. I. Oleynik, Physical Review B **94**, 125443 (2016).

[49] G. Domingo, R. S. Itoga, and C. R. Kannewurf, Physical Review **143**, 536 (1966).

[50] P. Manou, J. Kalomiros, A. Anagnostopoulos, and K. Kambas, Materials Research Bulletin **31**, 1407 (1996).